\documentclass[reprint,aps,prl,twocolumn,groupedaddress,superscriptaddress]{revtex4-2}
\usepackage{graphicx}
\usepackage{amsmath}
\usepackage{epstopdf}
\usepackage{float}
\usepackage{color}
\usepackage{appendix}
\usepackage{multirow}
\usepackage{booktabs}
\usepackage{longtable}
\usepackage{hyperref}

\begin{document}
\title{Orbital-resolved three-body recombination across a $p$-wave Feshbach resonance in ultracold $^6$Li}

\author{Shaokun Liu}
\thanks{These authors contributed equally to this work.}
\affiliation{School of Physics and Astronomy, Sun Yat-sen University, Zhuhai,  China 519082}

\author{Zhekang Xu}
\thanks{These authors contributed equally to this work.}
\affiliation{School of Physics and Astronomy, Sun Yat-sen University, Zhuhai,  China 519082}

\author{Shuai Peng}
\thanks{These authors contributed equally to this work.}
\affiliation{School of Physics and Astronomy, Sun Yat-sen University, Zhuhai,  China 519082}

\author{Sijia Peng}
\affiliation{School of Physics and Astronomy, Sun Yat-sen University, Zhuhai,  China 519082}

\author{Tangqian Shu}
\affiliation{School of Physics and Astronomy, Sun Yat-sen University, Zhuhai,  China 519082}

\author{Jiaming Li}
\email[]{lijiam29@mail.sysu.edu.cn}
\affiliation{School of Physics and Astronomy, Sun Yat-sen University, Zhuhai, China 519082}
\affiliation{Guangdong Provincial Key Laboratory of Quantum Metrology and Sensing,	Sun Yat-Sen University, Zhuhai, China 519082}
\affiliation{Center of Quantum Information Technology, Shenzhen Research Institute of Sun Yat-sen University, Shenzhen,  China 518087}
\affiliation{State Key Laboratory of Optoelectronic Materials and Technologies, Sun Yat-sen University, Guangzhou, China 510275}

\author{Le Luo}
\email[]{luole5@mail.sysu.edu.cn}
\affiliation{School of Physics and Astronomy, Sun Yat-sen University, Zhuhai,  China 519082}
\affiliation{Guangdong Provincial Key Laboratory of Quantum Metrology and Sensing,	Sun Yat-Sen University, Zhuhai, China 519082}
\affiliation{Center of Quantum Information Technology, Shenzhen Research Institute of Sun Yat-sen University, Shenzhen,  China 518087}
\affiliation{State Key Laboratory of Optoelectronic Materials and Technologies, Sun Yat-sen University, Guangzhou, China 510275}



\date{\today}

\begin{abstract}

	We report precision, orbital-resolved measurements of three-body recombination near the 159~G $p$-wave Feshbach resonance in an ultracold gas of $^{6}$Li atoms prepared in their lowest hyperfine state. Using a radio-frequency gated protocol that suppresses magnetic-field transients below the milligauss level, we resolve loss features associated with the $|m_\ell|=1$ and $m_\ell=0$ orbital projections. The measured three-body loss coefficient $L_3$ is well captured by a thermally averaged cascade-recombination model, enabling extraction of the resonance splitting $\delta B$ and effective-range parameter $k_e$. At the lowest temperature, we obtain $\delta B = 7.6(3)$~mG and $k_e = 0.151(6)\,a_0^{-1}$, both in quantitative agreement with coupled-channel theory.  These results establish orbital-resolved three-body spectroscopy as a precision probe of $p$-wave scattering and provide a benchmark for microscopic models of resonant few-body loss.

\end{abstract}
\maketitle


\textit{Introduction.}The ability to tune interatomic interactions via Feshbach resonances in ultracold atomic gases has enabled precise studies of few- and many-body quantum phenomena~\cite{Chin2010RMP82.1225,Bloch2008Rev.Mod.Phys.80.885-964}.
Among these, three-body recombination plays a central role by governing molecular formation~\cite{Li2018PRL120.193402}, determining the stability of strongly interacting gases~\cite{Du2009Phys.Rev.Lett.102.250402, Weber2003Phys.Rev.Lett.91.123201, Rem2013Phys.Rev.Lett.110.163202}, and providing access to emergent quantum phases~\cite{SaDeMelo1993Phys.Rev.Lett.71.3202-3205,Giorgini2008Rev.Mod.Phys.80.1215-1274}.
While $s$-wave three-body physics is now well established-with Efimov resonances and universal scaling laws characterized in multiple systems~\cite{Berninger2011PRL107.120401, Shotan2014Phys.Rev.Lett.113.053202, DIncao2018J.Phys.B:At.Mol.Opt.Phys.51.043001, Ji2022Phys.Rev.Lett.129.203402, Li2025Sci.China68.123011}-the analogous landscape for resonantly interacting $p$-wave systems remains much less complete.

$p$-wave interactions underlie a broad class of anisotropic and topological quantum phenomena, including chiral superfluidity and resonant few-body dynamics~\cite{Gaebler2007Phys.Rev.Lett.98.200403, Chevy2005PRA71.062710}.
Quantitative modeling of these systems requires precise knowledge of the $p$-wave scattering parameters: the orbital-resolved resonance splitting $\delta B$ and the effective-range coefficient $k_e$ that controls finite-energy behavior~\cite{Ticknor2004PRA69.042712, Luciuk2016Nat.Phys.12.599-605, Gerken2019RPA100.050701R, Peng2025PRL.135.133401}.
Ultracold $^6$Li near its 159~G $p$-wave Feshbach resonance provides an ideal platform for probing such anisotropic interactions, supported by accurately known molecular potentials and a well-studied resonance structure~\cite{Zhang2004PRA70.030702R, Cheng2005Phys.Rev.Lett.95.070404, Gurarie2007Ann.Phys.322.2-119, Buhler2014Nat.Commun.5.4504, Peng2024Phys.Rev.A110.L051301, Venu2023Nature613.262-267}.
However, a fully resolved and quantitatively bench-marked measurement of the $|m_\ell|=1$ and $m_\ell=0$ components has remained a significant experimental challenge, primarily due to the extreme sensitivity of the spectra to magnetic-field stability and transient asymmetries at the milligauss level.

Here, we overcome this challenge by performing orbital-resolved measurements of the three-body recombination rate for ultracold $^{6}$Li atoms in their lowest hyperfine state, $|1\rangle$, across the 159~G $p$-wave Feshbach resonance.
Using a radio frequency (RF)-gated state-preparation protocol that suppresses field-transient asymmetries, we clearly resolve the $|m_{\ell}|=1$ and $m_{\ell}=0$ features in the three-body loss coefficient $L_3$.
Modeling the spectra with a thermally averaged cascade recombination framework (atom $\rightarrow$ quasi-bound dimer $\rightarrow$ deeply bound dimer) enables quantitative extraction of the microscopic parameters $\delta B$ and $k_e$, which we compare directly to coupled-channel calculations.
Our results establish a precise experimental benchmark for $p$-wave few-body loss, clarify the role of orbital anisotropy and finite-temperature shifts, and map out the onset of unitarity-limited behavior in three-body recombination.

\begin{figure}[htbp]
	\centering
	\includegraphics[width=\columnwidth]{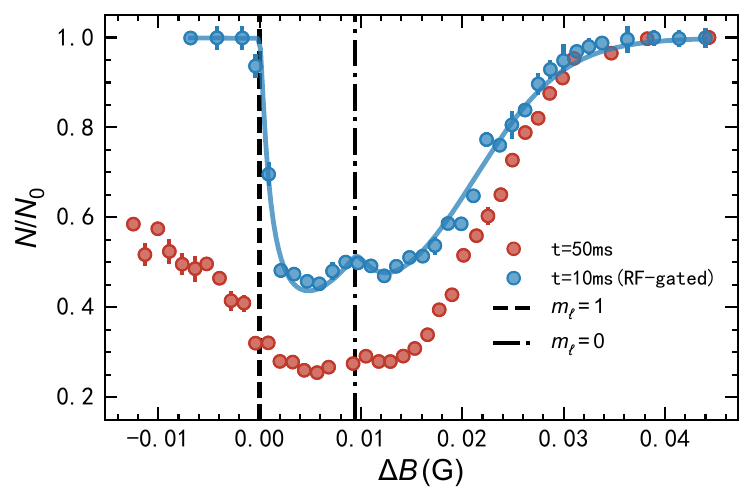}
	\caption{Comparison of atom-loss spectra near the 159~G $p$-wave Feshbach resonance obtained with and without the RF-gated protocol.
		Red points: spectra measured using the conventional method with a 50~ms holding time after the field step, showing a pronounced asymmetric low-detuning wing caused by eddy-current transients.
		Blue points: spectra taken using the RF-gated protocol with a 10~ms holding time, which suppresses field-transient asymmetries and restores the intrinsic symmetric line shape.
		Solid lines are fits to a model that accounts for three-body loss broadened by magnetic field noise, as detailed in Ref.~\cite{Peng2025PRL.135.133401}.
	}
	\label{fig1}
\end{figure}

\textit{Experiments.} We prepare an ultracold gas of $^{6}$Li in a crossed-beam optical dipole trap. Starting from a balanced $|1\rangle$ and $|2\rangle$ (two lowest hyperfine ground states) mixture at 320 G, forced evaporation sets the temperature $T$ (typically 0.16-1.2 $\mu$K, reduced temperature $T/T_F$>0.5 in these runs) and yields $N_0\approx 10^5$ atoms per spin \cite{Peng2025Phys.Rev.Res.7.023045}. Trap frequencies $\{\omega_x,\omega_y,\omega_z\}$ are calibrated by parametric heating. These frequencies determine the effective trap volume $V_{\rm eff} = \left(2\sqrt{3\pi k_B T / m \bar{\omega}^2}\right)^{3/2}$, which is used to extract $L_3$ from the atom loss data. The measured geometric mean frequency $\bar{\omega}=(\omega_x \omega_y \omega_z)^{1/3}$ and other detailed setup parameters are given in the Supplemental Material (SM). Magnetic fields are generated by low-noise Helmholtz coils driven by a bipolar current source\cite{Liu2023RSI94.053201}; the field is calibrated via RF spectroscopy and stabilized to the milligauss level at the atoms (residual rms noise $< 1$~mG) by feedforward technology\cite{Peng2025PRL.135.133401}. The eddy-current settling time following field steps is measured to be $\tau \approx 5$~ms\cite{Chen2021PRA103.063311}.

To obtain orbital-resolved measurements of three-body recombination rate, we must eliminate the systematic error induced by magnetic-field transients from the eddy-current effect, which makes the loss profile asymmetric. Here we employ an RF-gated state-preparation protocol. After evaporation, the sample is first ramped to 172~G on the high-field side of the 159~G $p$-wave resonance. Then a short resonant light pulse ($\sim\!100~\mu$s) removes the $|1\rangle$ atoms, yielding a pure $|2\rangle$ gas, which exhibits almost zero loss at the target fields during the duration of the measurement. The magnetic field is then stepped to $B = B_0 + \Delta B$ (here, $B_{0}= 159.215$ G, corresponding to the zero temperature resonance of $|m_{\ell}| = 1$ component) and held for 50~ms ($\approx 10\tau$) to allow eddy currents to decay, where the field is stabilized within 1~mG~\cite{Peng2025PRL.135.133401}. The interacting $|1\rangle$ population is subsequently created by a 2~ms RF Landau--Zener sweep with a range of $\pm 2.5\Omega$ and a Rabi frequency of $\Omega \approx 2\pi\times 2.0$~kHz. After holding at $B$ for a variable time $t$, the cloud is ramped back to 172~G and imaged after a 2~ms time-of-flight (TOF) to extract $N(t)$ and $T(t)$.

As shown in Fig.~\ref{fig1}, this protocol removes the asymmetric low-detuning wing observed in conventional field-switching methods and restores a symmetric spectrum in which the $|m_\ell|=1$ and $m_\ell=0$ features are clearly resolved. Because eddy-current transients are strongly suppressed, the loss signal acquired with a wait time of $t=10$ ms is comparable to that obtained after $t=50$ ms in the unsuppressed case. This advantage becomes particularly important at lower temperatures, where three-body loss is stronger and typically restricts hold times to below 100 ms.
This comparison highlights a key advantage of our protocol: by minimizing the wait time, it avoids the significant atom loss that occurs during the field settling in conventional methods, thereby reducing systematic uncertainty in the determination of $L_3$.

Then, the three-body recombination coefficient $L_3$ is obtained by fitting the measured $N(t)$ to
\begin{equation}
	\frac{dN}{dt} = -L_3 \frac{N^3}{ V_{\mathrm{eff}}^{2}},
	\label{eq:3body_loss}
\end{equation}
To account for the influence of $T(t)$ variations during the three-body process\cite{Top2021PRA104.043311, Peng2024Commun.Phys7.101}, we use
\begin{equation}
	N(t) = \left[N_0^{-2} + 2L_3 \int_0^{t} \frac{dt'}{V_{\mathrm{eff}}(t')^{2}}\right]^{-1/2},
	\label{eq:3body_solution}
\end{equation}
to fit the measured $N(t)$.
This expression incorporates the possible time dependence of $V_{\mathrm{eff}}(t)$ through the measured $T(t)$.
To verify the three-body nature of the observed decay, we measure (i) interacting scaling by varying the $\Delta B$, and (ii) two-body loss fits for comparison; all checks are summarized in SM.

\textit{Modeling.} Three-body recombination near a narrow $p$-wave Feshbach resonance can be described by a cascade (indirect) mechanism in which two atoms first associate into a quasi-bound $p$-wave molecule, which subsequently undergoes vibrational relaxation into a deeply bound dimer via a collision with a third atom. The population dynamics of the atomic and molecular densities, $n_{1}$ and $n_{D}$, follow the coupled rate equations
\begin{align}
	\frac{dn_{1}}{dt} & = 2\Gamma_{r} n_{D} - 2K_{M} n_{1}^{2} - K_{AD} n_{1} n_{D}, \\
	\frac{dn_{D}}{dt} & = -\Gamma_{r} n_{D} + K_{M} n_{1}^{2} - K_{AD} n_{1} n_{D},
\end{align}
where $K_{M}$ is the two-body association rate into the quasi-bound molecule, $K_{AD}$ is the atom-dimer relaxation rate, and $\Gamma_{r}(E) = 2\sqrt{m}\,E^{3/2}/(k_{e}\hbar^{2})$ is the energy-dependent resonance width of the quasi-bound molecular state.
Here, $m$ is the atomic mass, $E = \hbar^{2}k^{2}/m$ is the collision energy for a relative wave number $k$, and $\hbar$ is the reduced Planck constant. The parameter $k_{e}$ characterizes the energy dependence of the scattering via the effective-range expansion.

Under the assumption that the molecular population reaches a quasi-steady state much faster than the atomic population changes ($dn_D/dt \approx 0$), the molecular density is given by $n_D \approx K_M n_1^2 / (\Gamma_r + K_{AD} n_1)$. This assumption is valid when the dimer population equilibrates on a timescale much faster than the depletion of the atomic gas. Substituting this into the rate equation for $n_1$ yields an effective three-body loss process,
\begin{align}
	\frac{dn_1}{dt} = -\frac{3 K_{AD} K_M}{\Gamma_r  + K_{AD} n_1}\, n_1^3 \equiv -L_3 n_1^3.
\end{align}
The condition $\Gamma_r \gg K_{AD} n_1$ corresponds to the regime where the decay of the quasi-bound molecule back into two free atoms is much faster than its stabilization into a deep dimer via collision with a third atom. In this limit, the loss coefficient simplifies to
\begin{align}
	L_3 \approx \frac{3 K_{AD} K_M}{\Gamma_r }.
\end{align}

The two-body association rate follows a Breit--Wigner form~\cite{Gurarie2007Ann.Phys.322.2-119, Mathey2009Phys.Rev.A80.030702},
\begin{align}\label{kM(E)}
	K_M(E) = v \frac{\pi}{k^2}\
	\frac{3 \hbar^2 \Gamma_r^2(E)}{(E-E_r)^2 + \hbar^2 \Gamma_r^2(E)/4},
\end{align}
where $v$ is the relative velocity and the resonance energy (or binding energy) is
\begin{align}\label{E_r}
	E_{r} = \frac{\hbar^2}{m|V_p|k_e} \approx \frac{\hbar^{2}}{m V_{\mathrm{bg}}\Delta k_{e}} \Delta B,
\end{align}
where $V_p=V_{\mathrm{bg}}(1-\Delta/\Delta B)$ is the scattering volume. The approximation on the right is valid for small detunings $|\Delta B| \ll |\Delta|$. Here, $V_{\mathrm{bg}}$ and $\Delta$ are the background scattering volume and resonance width, respectively.

For a thermal gas, the association rate is thermally averaged over the Maxwell--Boltzmann distribution ($k_B$ denotes the Boltzmann constant),
\begin{align}
	\langle K_{M} \rangle_{T}
	= \frac{2}{\sqrt{\pi}(k_{B}T)^{3/2}}
	\int_{0}^{\infty} K_{M}(E)\, \sqrt{E}\, e^{-E/(k_{B}T)}\, dE,
\end{align}
and the corresponding thermally averaged loss rate is
\begin{align}
	\langle L_{3} \rangle_{T}
	= \frac{3 K_{AD} \langle K_{M} \rangle_{T}}{\Gamma_{r}}.
\end{align}

To account for the orbital anisotropy of the $p$-wave resonance, we treat the
$|m_{\ell}|=1$ and $m_{\ell}=0$ components independently.
Denoting $r$ as the fractional statistical weight of the $|m_{\ell}|=1$ component, the total thermally averaged rate is
\begin{equation}
	L_{3}^{\mathrm{total}}
	= r\, \langle L_{3}^{(|m_{\ell}|=1)} \rangle_{T}
	+ (1 - r)\, \langle L_{3}^{(m_{\ell}=0)} \rangle_{T}.
	\label{eq:fittingwithml}
\end{equation}

\begin{figure*}[htbp]
	\centering
	\includegraphics[width=1.0\textwidth]{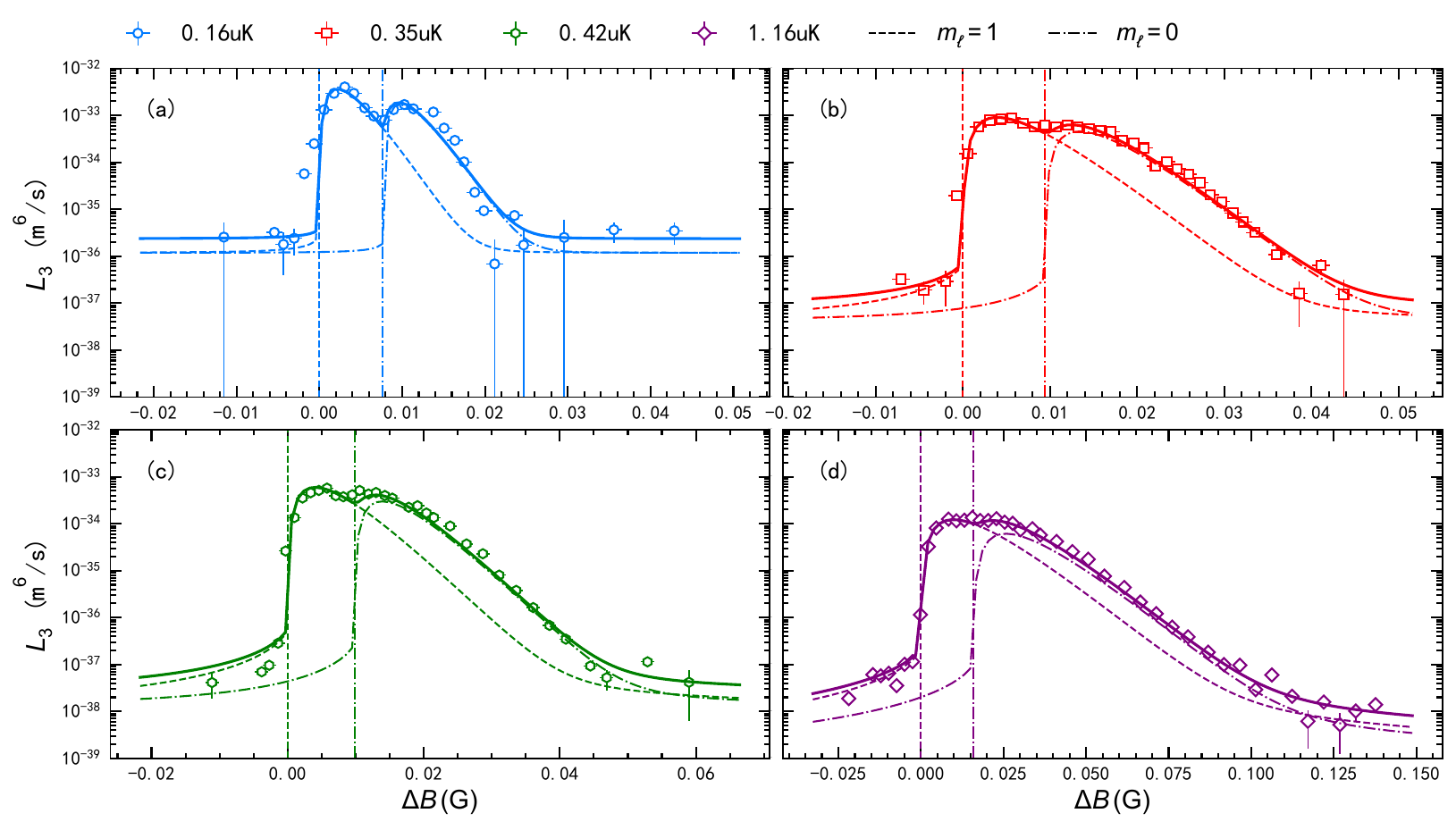}
	\caption{
		Three-body loss coefficient $L_{3}$ as a function of magnetic-field detuning $\Delta B$ at various temperatures $T$.
		Solid lines represent fits to Eq.~(\ref{eq:fittingwithml}).
		The $|m_{\ell}| = 1$ component and its corresponding zero-temperature resonance position are shown as dashed curves and vertical dashed lines, respectively, while the $m_{\ell} = 0$ component is shown as dash-dotted curves and lines.
		Vertical error bars indicate the $1\sigma$ statistical uncertainties obtained from fitting $N(t)$ with Eq.~(\ref{eq:3body_solution}).
		The raw data and model-fitted values are given in Appendix~\ref{app:A}, Tables~A1 -A4.
	}
	\label{fig:L3data}
\end{figure*}

\textit{Results.} Figure~\ref{fig:L3data} shows the measured $L_3$ as a function of $\Delta B$ near the 159~G $p$-wave Feshbach resonance for representative temperatures. The low-temperature data exhibit two clear maxima associated with the $\lvert m_\ell\rvert = 1$ and $m_\ell=0$ orbital projections. The clear resolution of these two features, separated by only $\sim 8$~mG, is a direct consequence of the magnetic field stability achieved with our RF-gated protocol, which suppresses transient effects that have obscured this splitting in previous measurements. The relative amplitudes and widths evolve systematically with temperature, demonstrating the interplay between orbital anisotropy and thermal averaging.

The spectra are analyzed using the cascade recombination model described above, thermally averaged according to Eq.~\ref{eq:fittingwithml} with a given offset, assuming $V_{\mathrm{bg}} = -4.236\times10^{4}\ a_{0}^{3}$ and $\Delta = -40.13~\mathrm{G}$ for all $m_{\ell}$ components~\cite{Fuchs2008PRA77.053616}. The model reproduces both the absolute magnitude and shape of $L_3$ across all measured temperatures with a single set of microscopic parameters: $K_{AD}$, $\delta B$, $r$, and  $k_e$. The extracted parameters are summarized in Table~\ref{tab:fit_parameters}.

\begin{table}[htbp]
	\centering
	\caption{Summary of the fitting parameters obtained from the fits shown in Fig. \ref{fig:L3data}. The final two row, $\delta B^u$ and $\bar\omega$, provides an estimate for the unitarity-limited broadening width and the used geometry mean trapping frequencies.}
	\label{tab:fit_parameters}
	\footnotesize
	\renewcommand{\arraystretch}{1.4} 
	\setlength{\tabcolsep}{2pt}
	\begin{tabular}{lcccc}
		\toprule
		\textbf{Parameter}                                    &
		\multicolumn{1}{c}{\textbf{0.16 $\boldsymbol{\mu}$K}} &
		\multicolumn{1}{c}{\textbf{0.35 $\boldsymbol{\mu}$K}} &
		\multicolumn{1}{c}{\textbf{0.42 $\boldsymbol{\mu}$K}} &
		\multicolumn{1}{c}{\textbf{1.16 $\boldsymbol{\mu}$K}}                                                 \\
		\midrule
		$\delta B$ (mG)                                       & 7.6(3)    & 9.4(3)    & 9.8(5)    & 15.9(9)   \\
		$K_{\mathrm{AD}}$ ($10^{-16}$ m$^{3}$/s)              & 1.68(21)  & 1.38(8)   & 1.19(8)   & 1.12(4)   \\
		$k_e$ ($a_0^{-1}$)                                    & 0.151(6)  & 0.125(3)  & 0.108(3)  & 0.095(2)  \\
		$r$                                                   & 0.68(5)   & 0.67      & 0.67      & 0.67      \\
		\hline
		$\delta B^{u}$ (mG)                                   & 1.1       & 2.3       & 2.8       & 7.7      \\
		$\bar{\omega}$ ($2 \pi \times \mathrm{Hz}$)           & 107(2)    & 156(2)    & 190(2)    & 351(2) \\
		\bottomrule
	\end{tabular}
\end{table}

At the lowest temperature, $T = 0.16~\mu$K, we obtain an orbital splitting of $\delta B = 7.6 \pm 0.3$~mG, in quantitative agreement with coupled-channel calculations of 7.58 mG based on high-precision $^{6}$Li $p$-wave interaction potentials~\cite{Zhang2022Chin.Phys.B31.063402, PrivateCommunication}. As the temperature increases, the apparent splitting shows a modest upward shift. This trend is consistent with finite-temperature and near-unitary effects, which shift the effective resonance position and are not explicitly included in the present model~\cite{Suno2003Phys.Rev.Lett.90.053202}.

The effective-range parameter at the lowest temperature, $k_e = 0.151 \pm 0.006~a_0^{-1}$, agrees well with the theoretical value $k_e^{\mathrm{th}} = 0.148~a_0^{-1}$ reported  previously~\cite{Fuchs2008PRA77.053616}. This agreement demonstrates the sensitivity of the three-body loss spectrum to the underlying $p$-wave scattering properties. The weak but systematic temperature dependence of the fitted $k_e$ likely reflects the onset of unitarity-limited behavior, where the collision cross-section approaches its maximum possible value and the simplified cascade model becomes less accurate.

The relative weight of the orbital components, $r \simeq 0.68$ for the $\lvert m_\ell\rvert = 1$ manifold, corresponds to the expected 2:1 statistical degeneracy in a three-dimensional gas~\cite{Kurlov2017PRA95.032710}. To improve fit stability and reduce correlations with the remaining microscopic parameters-especially in the regime where the two features overlap and near-unitary broadening becomes significant-we fix this ratio to $r = 0.67$ when analyzing the higher-temperature datasets.

The fitted atom-dimer relaxation rate $K_{AD}$ decreases gradually with $T$, from $1.7 \times 10^{-16}$~m$^{3}$/s at $T=0.16~\mu$K to $1.1 \times 10^{-16}$~m$^{3}$/s at $T=1.16~\mu$K. This trend is consistent with behavior observed near narrow $s$-wave resonances~\cite{Li2018PRL120.193402}, where van der Waals universality predicts the relaxation rate weakens as the collision energy increases.
For typical peak densities $n_1 \sim 10^{17}$~m$^{-3}$, the condition $\Gamma_r \gg K_{AD} n_1$ remains well satisfied across the entire temperature range, validating the use of the perturbative cascade limit underlying Eq.~\ref{eq:fittingwithml}.

The thermal averaging in our model naturally incorporates the temperature dependence of the loss spectra, including the systematic shift of the peak-loss position toward higher magnetic fields at elevated temperatures. As shown in Fig.~\ref{fig:L3data}, the zero-temperature resonance position $B_0$ (indicated by the dashed and dash-dotted lines) corresponds to the point of maximal curvature of $L_3$, rather than to the location of the peak itself—direct evidence of the strongly energy-dependent character of $p$-wave interactions.
Approximating the collision energy by its thermal mean, $\langle E\rangle = 3k_{B}T/2$, yields a temperature-dependent resonance position
\begin{equation}
	B_0(T) = B_0 + V_{\rm bg}\Delta k_e k_T^2,
	\label{eq:Bshift}
\end{equation}
where $k_T=\sqrt{3 m k_B T /(2\hbar^2)}$ is the thermal wavenumber.
Equation~(\ref{eq:Bshift}) reproduces the observed displacement of the peak-loss features in Fig.~\ref{fig:L3data} and highlights that, for narrow $p$-wave resonances, thermal shifts become comparable to the intrinsic linewidth. Consequently, these temperature-induced modifications must be included for quantitative, precision analysis of $L_3(B,T)$.

\begin{figure}[htbp]
	\centering
	\includegraphics[width=\columnwidth]{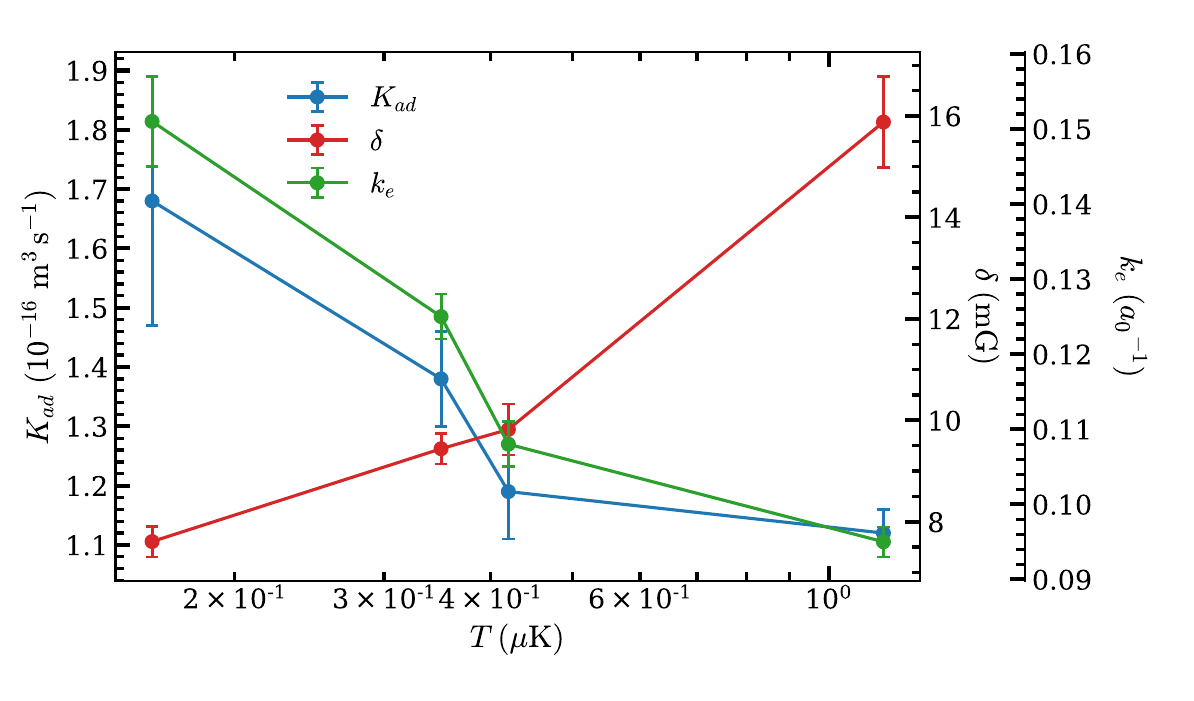}
	\caption{
		Extracted microscopic parameters $\delta B$ (red), $k_e$ (green), and $K_{AD}$ (blue) as a function of $T$. The parameters remain almost constant below $T \lesssim 0.4\,\mu$K, indicating faithful probing of zero-energy $p$-wave scattering. Deviations at higher $T$ reflect the onset of near-unitary broadening.
		Error bars indicate the $1\sigma$ fitting uncertainties.
	}
	\label{fig:parameters}
\end{figure}

Figure~\ref{fig:parameters} plots the weakly temperature dependence of $K_{AD}$, $\delta B$, and $k_e$.
The splitting $\delta B$ remains nearly constant for $T \lesssim 0.4~\mu$K, demonstrating that the low-temperature spectra faithfully probe the zero-energy scattering properties of the $p$-wave resonance. 
At higher temperatures, however, all extracted parameters exhibit systematic deviations from their low-$T$ limits. This marks the onset of the near-unitary regime, where the thermal collision energy becomes comparable to the intrinsic resonance width. In this regime, the characteristic unitary broadening $\delta B^{u}$ (Table~\ref{tab:fit_parameters}), which is equivalent to the thermal resonance shift in Eq.~\ref{eq:Bshift}, becomes non-negligible, and the simplified cascade model—lacking explicit unitarity and higher-order energy corrections—cannot fully reproduce the saturated line shape. As a result, the fits absorb these effects into the apparent values of $\delta B$, $k_e$, and $K_{AD}$.


\textit{Conclusion.} By developing a high-stability measurement protocol, we have successfully resolved the orbital components of three-body recombination near a narrow $p$-wave Feshbach resonance. This provides a direct experimental connection between orbital angular momentum projections and three-body recombination dynamics. In contrast to previous studies, which focused primarily on the strongly temperature-dependent regime, our measurements span the full accessible interaction range on both sides of the resonance. The data are analyzed using a complete, thermally averaged integral formulation of the cascade model. This treatment captures the correct energy dependence of elastic pair formation and the subsequent atom-dimer relaxation process, enabling quantitative fits across a wide range of detunings and temperatures.

The model reproduces the essential spectral features and allows precise extraction of the key microscopic parameters $\delta B$, $K_{AD}$, and $k_e$, with uncertainties consistent with state-of-the-art theoretical expectations. The small deviations observed at elevated temperatures highlight the onset of near-unitary broadening and motivate future extensions of the model to include explicit unitarity corrections and energy-dependent couplings. Further experimental refinements, such as measurements at lower densities to suppress many-body effects or with reduced trap heating, could enable even more stringent tests of $p$-wave scattering theory. The values reported here provide a precise benchmark for resonant $p$-wave few-body physics and help lay the groundwork for future studies of anisotropic pairing and topological superfluidity in ultracold fermionic systems.

\textit{Acknowledgements}
This work is supported by  NSFC under Grant No.12574302 and No.12174458. J.~Li received supports
from Fundamental Research Funds for Sun Yat-sen University 2023lgbj0 and 24xkjc015. L. Luo received supports from Shenzhen Science and Technology Program JCYJ20220818102003006.

%

\appendix
\renewcommand{\figurename}{SFig.}
\setcounter{figure}{0}
\renewcommand{\thetable}{S\arabic{table}}
\renewcommand{\theequation}{S.\arabic{equation}}
\setcounter{equation}{0}
\setcounter{table}{0}
\setcounter{section}{0}
\section*{Supplemental Material}

\section{More details about the time sequence}

\begin{figure}[htbp]
		\centering
		\includegraphics[width=1\columnwidth]{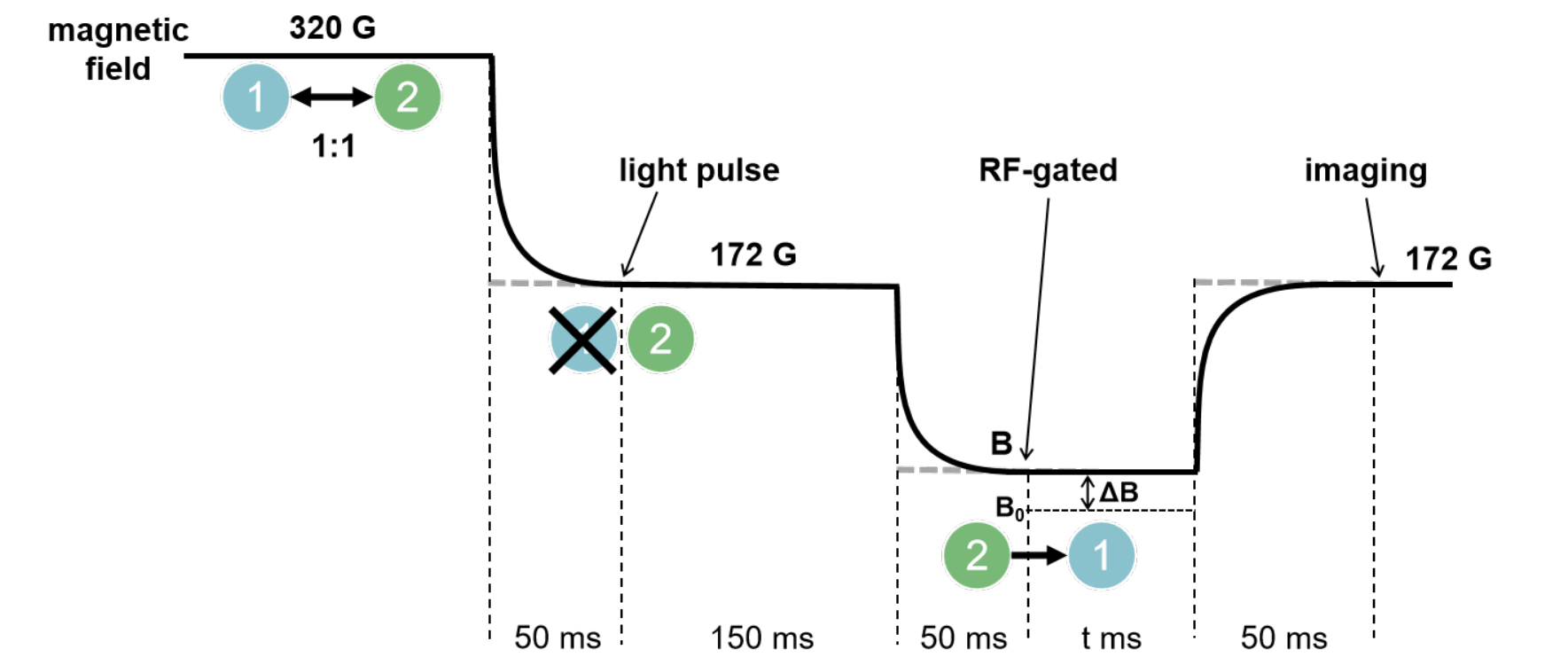}		
		\caption{Experimental time sequence of the magnetic field using the RF-gated protocol. }
		
		\label{fig:sequence}
\end{figure}
SFigure~\ref{fig:sequence} illustrates the complete magnetic-field timing sequence used in the RF-gated preparation protocol. The sequence is designed to suppress eddy-current--induced field transients to the milligauss level while minimizing the total wait time prior to initiating the loss dynamics.

The experiment begins with a mixture of atoms in the $|1\rangle$ and $|2\rangle$ hyperfine states confined in a crossed optical dipole trap at a magnetic field of 320~G. This field is chosen to ensure fast, stable evaporation and to suppress unwanted interaction effects during preparation.

The magnetic field is ramped to 172~G, well above the 159~G $p$-wave Feshbach resonance but low enough that the Zeeman splitting cleanly resolves the hyperfine states. At this field, a short resonant light pulse of $100 \mu$s selectively removes all atoms in the $|1\rangle$ state, resulting in a pure $|2\rangle$ sample. No measurable atom loss or heating is observed in this step.

The magnetic field is then stepped directly to the target interaction field,
\begin{equation}
	B = B_0 + \Delta B,
\end{equation}
where $B_0$ is the center of the $p$-wave resonance. After the step, the field is held for 50 ms---approximately ten times the decay constant of the dominant eddy-current mode in the aluminum vacuum chamber. This wait ensures that residual field transients decay to below $\sim$1\,mG, which is essential for obtaining symmetric and reproducible loss spectra.

Once the field has stabilized, a 2 ms RF Landau--Zener sweep transfers a controlled fraction of atoms from $|2\rangle$ to the interacting $|1\rangle$ state. The sweep employs a bandwidth of approximately $5\Omega$, with a Rabi frequency of $\Omega \approx 2\pi \times 2$~kHz, ensuring adiabatic population transfer. The end of the sweep defines $t = 0$ for the subsequent atom-loss dynamics.

The cloud is then held at the interaction field $B$ for a variable time $t$ (typically 5--200~ms), during which three-body recombination is measured. The RF-gated protocol ensures that no significant atom loss occurs prior to $t=0$, in contrast to conventional field-switching methods.

After the hold time, the field is ramped back to 172~G, where standard absorption imaging is performed after a 2 ms time-of-flight. Atom number $N(t)$ and temperature $T(t)$ are extracted by fitting the optical-density distribution.

\section{Time traces of $N(t)$ and $T(t)$ at selected $\Delta B$}

\begin{figure*}[t]
	\centering
	\includegraphics[width=1\textwidth]{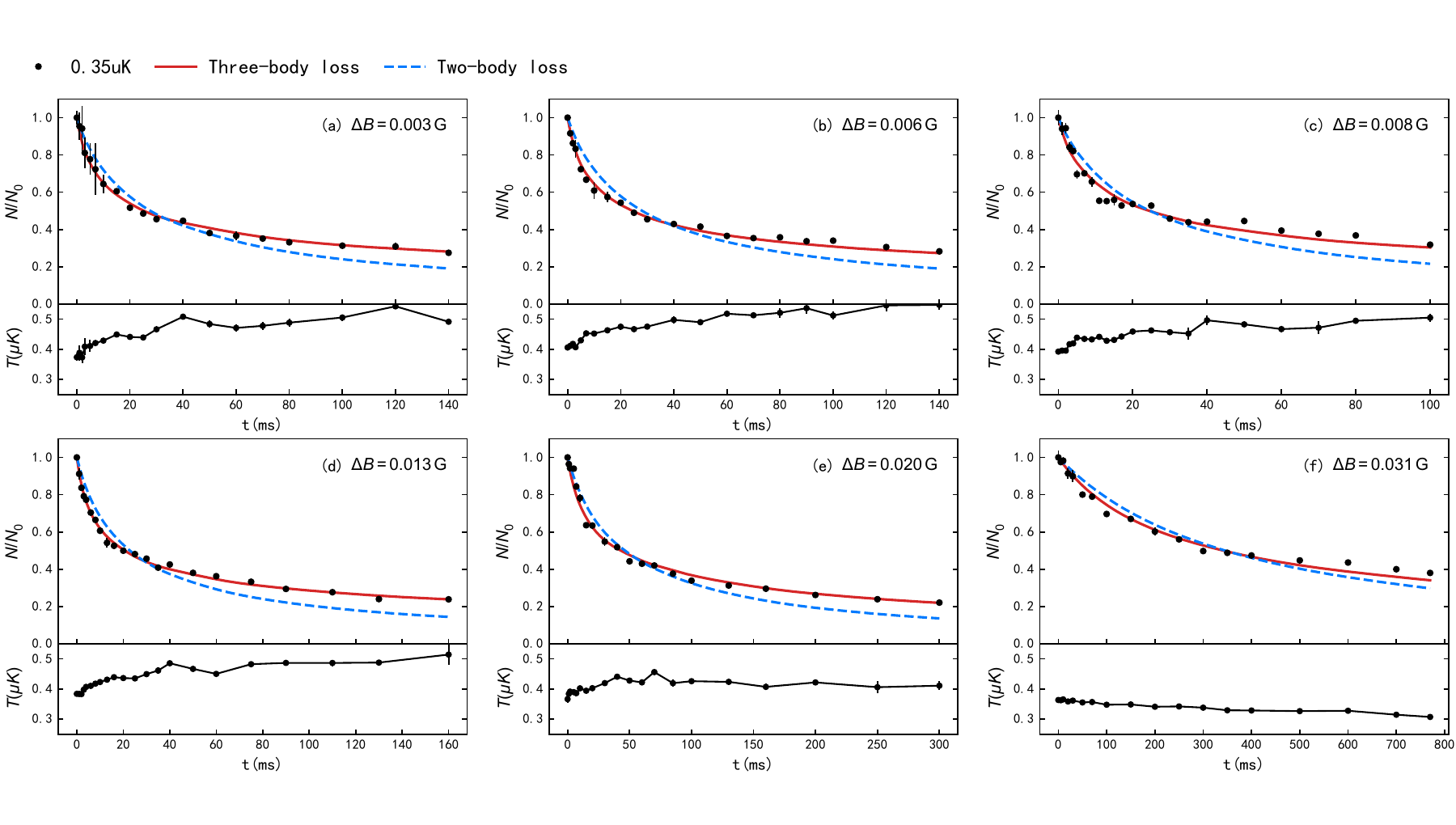}
	\caption{
		Time evolution of the atom number $N(t)$ (upper panels) and temperature $T(t)$ (lower panels) at selected magnetic field detunings $\Delta B$ corresponding to the features in Fig.~\ref{fig:L3data}(b).
		Red solid curves represent fits using Eq.~(\ref{eq:3body_loss}), while blue dashed curves show two-body fits for comparison. In the fitting procedure, $V_{\mathrm{eff}}(t)$ is constructed from $T(t)$ and linearly interpolated in time to evaluate the integral $\int_0^{t} 1/V_{\mathrm{eff}}(t')^{2}dt'$.
		The data points represent averages over 3 repeated measurements, with error bars indicating one standard deviation of the mean.
	}
	\label{fig2}
\end{figure*}

SFigure~\ref{fig2} shows representative time traces of $N(t)$ and $T(t)$ at several detunings $\Delta B$ for the $T = 0.35~\mu$K data set discussed in Fig.~2(b) of the main text.
The red solid curves are fits obtained using the full three-body model of Eq.~(2), whereas the blue dashed curves correspond to an analytic two-body decay model.
The excellent agreement between the experimental data and the three-body model across the entire range of $\Delta B$ confirms that three-body recombination constitutes the dominant loss mechanism, even in the region where the $|m_\ell|=1$ and $m_\ell=0$ components overlap.
This analysis also demonstrates that the heating dynamics encoded in $T(t)$ are essential for obtaining self-consistent fits, particularly near the peak-loss regions.

\section{Comparison with measurements and models in previous work}

A central question in the study of $L_3$ concerns its scaling laws and threshold behavior.
In Ref.~\cite{Yoshida2018PRL120.133401SM}, the scaling law for the three-body loss rate on the BCS side of the $|1\rangle$-$|1\rangle$ $p$-wave Feshbach resonance was experimentally verified to follow $L_3 \propto V_p^{8/3}$, consistent with the theoretical prediction based on dimensional analysis in Ref.~\cite{Suno2003Phys.Rev.Lett.90.053202SM}. Their measurement showed that this scaling holds only in the regime relatively far from resonance, where $V_p$ remains finite and $L_3$ is well below $1\times10^{-37} \mathrm{m}^6 \mathrm{s}^{-1}$. Such a low loss rate requires an exceptionally long trap lifetimes (over a 100 seconds), which poses a significant experimental challenge for most groups.

In our experiment, the minimum observed $L_3$ is already larger than $1\times10^{-37} \mathrm{m}^6 \mathrm{s}^{-1}$, limited by the 33 seconds lifetime of our trap. This indicates that we have not accessed the weakly interacting regime explored in Ref.~\cite{Yoshida2018PRL120.133401SM}, and consequently we cannot experimentally confirm the $V_p^{8/3}$ scaling law. Instead, our measurements probe a more strongly interacting regime, closer to resonance, where the simple power-law dependence breaks down. In the stronger interacting regime,our theoretical model likewise does not predict a universal scaling law for the three-body loss process in this strongly interacting limit.

A useful comparison can be made with earlier theoretical work that also employed a cascade description of $p$-wave three-body loss. In that approach, the initial two-body step was modeled by adding an imaginary component to the scattering volume in the coefficient $K_{M}$ in order to account for inelastic relaxation, which led to a sharp peak feature near resonance~\cite{Waseem2018Phys.Rev.A98.020702SM}. In their subsequent study~\cite{Waseem2019PRA99.052704SM}, however, this assumption was no longer adopted. Instead, the two-body collision was treated as purely elastic, and three-body loss  arising solely from atom-dimer relaxation following elastic pair formation, consistent with the absence of any energetically allowed two-body decay channel for the $|1\rangle$-$|1\rangle$ $p$-wave interaction.

Although both our work and Ref.~\cite{Waseem2019PRA99.052704SM} use this elastic cascade framework, the treatments of the atom-dimer association rate $K_M$ differ. In their work, $K_M$ was approximated analytically, enabling a closed-form expression but limiting the applicability of the model to the BCS side where the approximation remains accurate. In contrast, our analysis evaluates $K_M$ in its full thermally averaged integral form without approximation. This complete treatment captures the correct energy dependence across the entire interaction range and therefore allows us to model both the BCS and BEC regimes on equal footing.

With these considerations, the smoother resonance profiles observed in the experiment arise naturally in our model, whereas introducing an imaginary term in $K_M$ produces an unrealistically sharp peak, as seen in Fig.~1 of Ref.~\cite{Waseem2018Phys.Rev.A98.020702SM}. The consistently obtained behavior supports the interpretation that the observed losses originate from atom-dimer relaxation following elastic pair formation, rather than from intrinsic two-body decay.
\section{Trapping frequencies and temperature}

The trapping potential is provided by a crossed ODT plus a weak magnetic confinement. The ODT trap frequencies $\omega_i$ are calibrated from parametric heating at each working point, while the magnetic trap frequencies are measured once from center-of-mass oscillations and kept fixed at $\omega_{mx}=\omega_{my}=2\pi\times11.7(3) \mathrm{Hz}$ and $\omega_{mz}=2\pi\times16.6(4) \mathrm{Hz}$. The effective confinement frequencies used in the analysis are obtained by combining the ODT and magnetic contributions, $\tilde{\omega}_x^2=\omega_x^2+\omega_{mx}^2$, $\tilde{\omega}_y^2=\omega_y^2+\omega_{my}^2$, and $\tilde{\omega}_z^2=\omega_z^2-\omega_{mz}^2$. The measured ODT frequencies, together with the corresponding temperatures, average atom numbers, and degeneracy parameters $T/T_F$, are summarized in Table~\ref{tab:trap_parameters}.

\begin{table}[htbp]
	\centering
	\scriptsize
	\setlength{\tabcolsep}{6pt}
	\renewcommand{\arraystretch}{1.0}
	\begin{tabular}{c|c|c|c|c|c}
		\hline\hline
		$T$      & $\omega_x$            & $\omega_y$            & $\omega_z$            & $N_{\mathrm{avg}}$ & $T/T_F$ \\
		($\mu$K) & $(2\pi\,\mathrm{Hz})$ & $(2\pi\,\mathrm{Hz})$ & $(2\pi\,\mathrm{Hz})$ &                    &         \\
		\hline
		0.16     & 218.9(13)             & 22.9(11)              & 217.8(13)             & $7.9\times10^4$    & 0.59    \\
		0.35     & 346.0(21)             & 29.3(13)              & 344.3(21)             & $2.5\times10^5$    & 0.52    \\
		0.42     & 437.7(26)             & 34.0(10)              & 435.6(26)             & $2.9\times10^5$    & 0.51    \\
		1.16     & 773.8(46)             & 71.6(07)              & 769.9(46)             & $4.3\times10^5$    & 0.62    \\
		\hline\hline
	\end{tabular}
	\caption{Trap frequencies, average atom number $N_{\mathrm{avg}}$, and
		degeneracy parameter $T/T_F$ for the four temperatures used in this work.}
	\label{tab:trap_parameters}
\end{table}

\onecolumngrid

\section{Raw data and model-fitted values of $L_3$}
\refstepcounter{section}\label{app:A}

\setcounter{table}{0}
\renewcommand{\thetable}{A\arabic{table}}

\begin{table*}[htbp]
	\renewcommand{\arraystretch}{1.25}
	\setlength{\tabcolsep}{20pt}
	\centering
	\footnotesize

	\caption{Experimental three-body loss rate $L_3^{\mathrm{exp}}$ as a function of magnetic field $\Delta B$ at $T=0.16~\mu\mathrm{K}$,together with theoretical results $L_3^{\mathrm{th}}$ obtained from the model fit,and its two partial contributions corresponding to $|m_\ell|=1$ and $m_\ell=0$ channels.}
	\label{tab:S1}
	\begin{tabular}{|c|c|c|c|c|}
		\hline
		$\Delta B$ (G) & $L_3^{\rm exp}$ (m$^6$/s)                 & $L_3^{\rm th}$ (m$^6$/s) & $L_3^{\rm th}(|m_{\ell}|=1)$ (m$^6$/s) & $L_3^{\rm th}(m_{\ell}=0)$ (m$^6$/s) \\
		\hline
		-0.0115        & $2.55\times10^{-36}\pm2.66\times10^{-36}$ & $2.43\times10^{-36}$     & $1.24\times10^{-36}$                   & $1.19\times10^{-36}$                 \\
		-0.0054        & $3.23\times10^{-36}\pm6.25\times10^{-37}$ & $2.54\times10^{-36}$     & $1.34\times10^{-36}$                   & $1.20\times10^{-36}$                 \\
		-0.0043        & $1.79\times10^{-36}\pm1.39\times10^{-36}$ & $2.60\times10^{-36}$     & $1.39\times10^{-36}$                   & $1.20\times10^{-36}$                 \\
		-0.0030        & $2.41\times10^{-36}\pm1.36\times10^{-36}$ & $2.69\times10^{-36}$     & $1.48\times10^{-36}$                   & $1.21\times10^{-36}$                 \\
		-0.0018        & $5.72\times10^{-35}\pm7.52\times10^{-36}$ & $2.85\times10^{-36}$     & $1.64\times10^{-36}$                   & $1.21\times10^{-36}$                 \\
		-0.0006        & $2.50\times10^{-34}\pm2.74\times10^{-35}$ & $3.25\times10^{-36}$     & $2.03\times10^{-36}$                   & $1.22\times10^{-36}$                 \\
		0.0006         & $1.32\times10^{-33}\pm1.21\times10^{-34}$ & $1.40\times10^{-33}$     & $1.40\times10^{-33}$                   & $1.23\times10^{-36}$                 \\
		0.0018         & $2.95\times10^{-33}\pm1.80\times10^{-34}$ & $3.56\times10^{-33}$     & $3.56\times10^{-33}$                   & $1.25\times10^{-36}$                 \\
		0.0031         & $4.03\times10^{-33}\pm2.77\times10^{-34}$ & $3.29\times10^{-33}$     & $3.29\times10^{-33}$                   & $1.27\times10^{-36}$                 \\
		0.0042         & $2.95\times10^{-33}\pm2.75\times10^{-34}$ & $2.45\times10^{-33}$     & $2.45\times10^{-33}$                   & $1.30\times10^{-36}$                 \\
		0.0055         & $1.45\times10^{-33}\pm7.68\times10^{-35}$ & $1.50\times10^{-33}$     & $1.50\times10^{-33}$                   & $1.37\times10^{-36}$                 \\
		0.0066         & $9.70\times10^{-34}\pm4.57\times10^{-35}$ & $9.30\times10^{-34}$     & $9.28\times10^{-34}$                   & $1.49\times10^{-36}$                 \\
		0.0079         & $7.94\times10^{-34}\pm3.73\times10^{-35}$ & $8.06\times10^{-34}$     & $4.96\times10^{-34}$                   & $3.10\times10^{-34}$                 \\
		0.0090         & $1.33\times10^{-33}\pm6.48\times10^{-35}$ & $1.72\times10^{-33}$     & $2.82\times10^{-34}$                   & $1.44\times10^{-33}$                 \\
		0.0103         & $1.70\times10^{-33}\pm5.30\times10^{-35}$ & $1.76\times10^{-33}$     & $1.42\times10^{-34}$                   & $1.62\times10^{-33}$                 \\
		0.0114         & $1.38\times10^{-33}\pm3.82\times10^{-35}$ & $1.37\times10^{-33}$     & $7.73\times10^{-35}$                   & $1.29\times10^{-33}$                 \\
		0.0138         & $1.18\times10^{-33}\pm5.64\times10^{-35}$ & $5.39\times10^{-34}$     & $2.09\times10^{-35}$                   & $5.18\times10^{-34}$                 \\
		0.0151         & $5.36\times10^{-34}\pm1.86\times10^{-35}$ & $2.91\times10^{-34}$     & $1.04\times10^{-35}$                   & $2.81\times10^{-34}$                 \\
		0.0164         & $2.94\times10^{-34}\pm1.10\times10^{-35}$ & $1.51\times10^{-34}$     & $5.46\times10^{-36}$                   & $1.45\times10^{-34}$                 \\
		0.0175         & $1.02\times10^{-34}\pm4.18\times10^{-36}$ & $8.63\times10^{-35}$     & $3.48\times10^{-36}$                   & $8.28\times10^{-35}$                 \\
		0.0188         & $2.30\times10^{-35}\pm2.22\times10^{-36}$ & $4.33\times10^{-35}$     & $2.27\times10^{-36}$                   & $4.10\times10^{-35}$                 \\
		0.0199         & $9.31\times10^{-36}\pm1.95\times10^{-36}$ & $2.45\times10^{-35}$     & $1.78\times10^{-36}$                   & $2.27\times10^{-35}$                 \\
		0.0212         & $6.86\times10^{-37}\pm1.63\times10^{-36}$ & $1.30\times10^{-35}$     & $1.49\times10^{-36}$                   & $1.15\times10^{-35}$                 \\
		0.0236         & $7.37\times10^{-36}\pm2.29\times10^{-36}$ & $5.01\times10^{-36}$     & $1.29\times10^{-36}$                   & $3.72\times10^{-36}$                 \\
		0.0247         & $1.73\times10^{-36}\pm2.22\times10^{-36}$ & $3.78\times10^{-36}$     & $1.26\times10^{-36}$                   & $2.53\times10^{-36}$                 \\
		0.0296         & $2.51\times10^{-36}\pm3.44\times10^{-36}$ & $2.48\times10^{-36}$     & $1.21\times10^{-36}$                   & $1.27\times10^{-36}$                 \\
		0.0356         & $3.65\times10^{-36}\pm1.70\times10^{-36}$ & $2.39\times10^{-36}$     & $1.20\times10^{-36}$                   & $1.20\times10^{-36}$                 \\
		0.0429         & $3.48\times10^{-36}\pm1.75\times10^{-36}$ & $2.38\times10^{-36}$     & $1.19\times10^{-36}$                   & $1.19\times10^{-36}$                 \\
		\hline
	\end{tabular}
\end{table*}

\begin{table*}[htbp]
	\renewcommand{\arraystretch}{1.25}
	\setlength{\tabcolsep}{20pt}
	\centering
	\footnotesize

	\caption{Experimental three-body loss rate $L_3^{\mathrm{exp}}$ as a function of magnetic field $\Delta B$ at $T=0.35~\mu\mathrm{K}$,together with theoretical results $L_3^{\mathrm{th}}$ obtained from the model fit,and its two partial contributions corresponding to $|m_\ell|=1$ and $m_\ell=0$ channels.}
	\label{tab:S2}
	\begin{tabular}{|c|c|c|c|c|}
		\hline
		$B$ (G) & $L_3^{\rm exp}$ (m$^6$/s)                 & $L_3^{\rm th}$ (m$^6$/s) & $L_3^{\rm th}(|m_{\ell}|=1)$ (m$^6$/s) & $L_3^{\rm th}(m_{\ell}=0)$ (m$^6$/s) \\
		\hline
		-0.0071 & $3.23\times10^{-37}\pm6.32\times10^{-38}$ & $2.03\times10^{-37}$     & $1.44\times10^{-37}$                   & $5.85\times10^{-38}$                 \\
		-0.0045 & $1.89\times10^{-37}\pm9.64\times10^{-38}$ & $2.63\times10^{-37}$     & $1.99\times10^{-37}$                   & $6.34\times10^{-38}$                 \\
		-0.0020 & $2.91\times10^{-37}\pm2.06\times10^{-37}$ & $3.89\times10^{-37}$     & $3.19\times10^{-37}$                   & $7.01\times10^{-38}$                 \\
		-0.0007 & $1.95\times10^{-35}\pm1.63\times10^{-36}$ & $5.49\times10^{-37}$     & $4.74\times10^{-37}$                   & $7.49\times10^{-38}$                 \\
		0.0006  & $1.54\times10^{-34}\pm1.21\times10^{-35}$ & $1.97\times10^{-34}$     & $1.96\times10^{-34}$                   & $8.09\times10^{-38}$                 \\
		0.0018  & $5.74\times10^{-34}\pm2.96\times10^{-35}$ & $6.23\times10^{-34}$     & $6.23\times10^{-34}$                   & $8.79\times10^{-38}$                 \\
		0.0031  & $7.87\times10^{-34}\pm3.34\times10^{-35}$ & $8.67\times10^{-34}$     & $8.67\times10^{-34}$                   & $9.80\times10^{-38}$                 \\
		0.0044  & $8.34\times10^{-34}\pm2.50\times10^{-35}$ & $8.97\times10^{-34}$     & $8.97\times10^{-34}$                   & $1.12\times10^{-37}$                 \\
		0.0056  & $8.79\times10^{-34}\pm5.94\times10^{-35}$ & $8.14\times10^{-34}$     & $8.14\times10^{-34}$                   & $1.30\times10^{-37}$                 \\
		0.0069  & $6.84\times10^{-34}\pm3.82\times10^{-35}$ & $6.78\times10^{-34}$     & $6.78\times10^{-34}$                   & $1.62\times10^{-37}$                 \\
		0.0082  & $5.74\times10^{-34}\pm3.46\times10^{-35}$ & $5.36\times10^{-34}$     & $5.35\times10^{-34}$                   & $2.18\times10^{-37}$                 \\
		0.0094  & $6.15\times10^{-34}\pm3.38\times10^{-35}$ & $4.35\times10^{-34}$     & $4.16\times10^{-34}$                   & $1.96\times10^{-35}$                 \\
		0.0107  & $5.71\times10^{-34}\pm2.00\times10^{-35}$ & $5.32\times10^{-34}$     & $3.07\times10^{-34}$                   & $2.25\times10^{-34}$                 \\
		0.0120  & $6.26\times10^{-34}\pm1.50\times10^{-35}$ & $6.25\times10^{-34}$     & $2.22\times10^{-34}$                   & $4.03\times10^{-34}$                 \\
		0.0132  & $5.55\times10^{-34}\pm2.04\times10^{-35}$ & $6.14\times10^{-34}$     & $1.62\times10^{-34}$                   & $4.52\times10^{-34}$                 \\
		0.0145  & $5.17\times10^{-34}\pm1.93\times10^{-35}$ & $5.44\times10^{-34}$     & $1.14\times10^{-34}$                   & $4.30\times10^{-34}$                 \\
		0.0158  & $4.76\times10^{-34}\pm2.96\times10^{-35}$ & $4.49\times10^{-34}$     & $7.90\times10^{-35}$                   & $3.70\times10^{-34}$                 \\
		0.0170  & $4.50\times10^{-34}\pm3.54\times10^{-35}$ & $3.59\times10^{-34}$     & $5.57\times10^{-35}$                   & $3.03\times10^{-34}$                 \\
		0.0183  & $2.90\times10^{-34}\pm1.93\times10^{-35}$ & $2.72\times10^{-34}$     & $3.79\times10^{-35}$                   & $2.34\times10^{-34}$                 \\
		0.0196  & $2.53\times10^{-34}\pm1.77\times10^{-35}$ & $2.00\times10^{-34}$     & $2.56\times10^{-35}$                   & $1.75\times10^{-34}$                 \\
		0.0208  & $2.02\times10^{-34}\pm1.05\times10^{-35}$ & $1.49\times10^{-34}$     & $1.79\times10^{-35}$                   & $1.31\times10^{-34}$                 \\
		0.0221  & $8.34\times10^{-35}\pm3.60\times10^{-36}$ & $1.06\times10^{-34}$     & $1.20\times10^{-35}$                   & $9.39\times10^{-35}$                 \\
		0.0234  & $1.04\times10^{-34}\pm8.67\times10^{-36}$ & $7.43\times10^{-35}$     & $8.00\times10^{-36}$                   & $6.63\times10^{-35}$                 \\
		0.0246  & $7.22\times10^{-35}\pm2.85\times10^{-36}$ & $5.29\times10^{-35}$     & $5.49\times10^{-36}$                   & $4.74\times10^{-35}$                 \\
		0.0259  & $5.59\times10^{-35}\pm3.30\times10^{-36}$ & $3.63\times10^{-35}$     & $3.65\times10^{-36}$                   & $3.26\times10^{-35}$                 \\
		0.0272  & $3.66\times10^{-35}\pm1.84\times10^{-36}$ & $2.48\times10^{-35}$     & $2.44\times10^{-36}$                   & $2.23\times10^{-35}$                 \\
		0.0284  & $2.04\times10^{-35}\pm5.62\times10^{-37}$ & $1.73\times10^{-35}$     & $1.69\times10^{-36}$                   & $1.57\times10^{-35}$                 \\
		0.0297  & $1.45\times10^{-35}\pm6.36\times10^{-37}$ & $1.17\times10^{-35}$     & $1.14\times10^{-36}$                   & $1.06\times10^{-35}$                 \\
		0.0310  & $8.27\times10^{-36}\pm4.03\times10^{-37}$ & $7.88\times10^{-36}$     & $7.71\times10^{-37}$                   & $7.11\times10^{-36}$                 \\
		0.0322  & $5.41\times10^{-36}\pm2.37\times10^{-37}$ & $5.44\times10^{-36}$     & $5.44\times10^{-37}$                   & $4.89\times10^{-36}$                 \\
		0.0335  & $3.20\times10^{-36}\pm1.63\times10^{-37}$ & $3.65\times10^{-36}$     & $3.81\times10^{-37}$                   & $3.27\times10^{-36}$                 \\
		0.0360  & $1.08\times10^{-36}\pm8.28\times10^{-38}$ & $1.72\times10^{-36}$     & $2.09\times10^{-37}$                   & $1.51\times10^{-36}$                 \\
		0.0386  & $1.61\times10^{-37}\pm1.30\times10^{-37}$ & $8.16\times10^{-37}$     & $1.27\times10^{-37}$                   & $6.89\times10^{-37}$                 \\
		0.0411  & $6.40\times10^{-37}\pm2.40\times10^{-37}$ & $4.29\times10^{-37}$     & $9.16\times10^{-38}$                   & $3.37\times10^{-37}$                 \\
		0.0437  & $1.54\times10^{-37}\pm1.61\times10^{-37}$ & $2.51\times10^{-37}$     & $7.38\times10^{-38}$                   & $1.77\times10^{-37}$                 \\
		\hline
	\end{tabular}
\end{table*}

\begin{table*}[htbp]
	\renewcommand{\arraystretch}{1.25}
	\setlength{\tabcolsep}{20pt}
	\centering
	\footnotesize

	\caption{Experimental three-body loss rate $L_3^{\mathrm{exp}}$ as a function of magnetic field $\Delta B$ at $T=0.42~\mu\mathrm{K}$,together with theoretical results $L_3^{\mathrm{th}}$ obtained from the model fit,and its two partial contributions corresponding to $|m_\ell|=1$ and $m_\ell=0$ channels.}
	\label{tab:S3}
	\begin{tabular}{|c|c|c|c|c|}
		\hline
		$B$ (G) & $L_3^{\rm exp}$ (m$^6$/s)                 & $L_3^{\rm th}$ (m$^6$/s) & $L_3^{\rm th}(|m_{\ell}|=1)$ (m$^6$/s) & $L_3^{\rm th}(m_{\ell}=0)$ (m$^6$/s) \\
		\hline
		-0.0111 & $4.12\times10^{-38}\pm2.23\times10^{-38}$ & $9.12\times10^{-38}$     & $6.72\times10^{-38}$                   & $2.39\times10^{-38}$                 \\
		-0.0038 & $6.98\times10^{-38}\pm1.15\times10^{-38}$ & $2.02\times10^{-37}$     & $1.69\times10^{-37}$                   & $3.34\times10^{-38}$                 \\
		-0.0027 & $9.63\times10^{-38}\pm1.02\times10^{-38}$ & $2.48\times10^{-37}$     & $2.12\times10^{-37}$                   & $3.58\times10^{-38}$                 \\
		-0.0014 & $2.84\times10^{-37}\pm3.07\times10^{-38}$ & $3.37\times10^{-37}$     & $2.98\times10^{-37}$                   & $3.92\times10^{-38}$                 \\
		-0.0003 & $2.62\times10^{-35}\pm4.79\times10^{-36}$ & $4.87\times10^{-37}$     & $4.44\times10^{-37}$                   & $4.27\times10^{-38}$                 \\
		0.0010  & $1.35\times10^{-34}\pm1.33\times10^{-35}$ & $2.20\times10^{-34}$     & $2.19\times10^{-34}$                   & $4.79\times10^{-38}$                 \\
		0.0022  & $3.53\times10^{-34}\pm2.98\times10^{-35}$ & $4.66\times10^{-34}$     & $4.66\times10^{-34}$                   & $5.39\times10^{-38}$                 \\
		0.0034  & $4.57\times10^{-34}\pm4.65\times10^{-35}$ & $5.79\times10^{-34}$     & $5.79\times10^{-34}$                   & $6.18\times10^{-38}$                 \\
		0.0046  & $5.11\times10^{-34}\pm3.01\times10^{-35}$ & $5.84\times10^{-34}$     & $5.84\times10^{-34}$                   & $7.25\times10^{-38}$                 \\
		0.0058  & $5.73\times10^{-34}\pm2.49\times10^{-35}$ & $5.33\times10^{-34}$     & $5.32\times10^{-34}$                   & $8.71\times10^{-38}$                 \\
		0.0071  & $3.94\times10^{-34}\pm8.86\times10^{-36}$ & $4.47\times10^{-34}$     & $4.46\times10^{-34}$                   & $1.12\times10^{-37}$                 \\
		0.0082  & $3.75\times10^{-34}\pm8.46\times10^{-36}$ & $3.69\times10^{-34}$     & $3.69\times10^{-34}$                   & $1.46\times10^{-37}$                 \\
		0.0095  & $4.12\times10^{-34}\pm1.45\times10^{-35}$ & $2.85\times10^{-34}$     & $2.84\times10^{-34}$                   & $6.58\times10^{-37}$                 \\
		0.0106  & $5.12\times10^{-34}\pm3.27\times10^{-35}$ & $3.06\times10^{-34}$     & $2.24\times10^{-34}$                   & $8.22\times10^{-35}$                 \\
		0.0119  & $4.29\times10^{-34}\pm1.02\times10^{-35}$ & $3.88\times10^{-34}$     & $1.65\times10^{-34}$                   & $2.23\times10^{-34}$                 \\
		0.0130  & $4.63\times10^{-34}\pm1.04\times10^{-35}$ & $4.10\times10^{-34}$     & $1.25\times10^{-34}$                   & $2.85\times10^{-34}$                 \\
		0.0143  & $3.99\times10^{-34}\pm1.31\times10^{-35}$ & $3.84\times10^{-34}$     & $8.96\times10^{-35}$                   & $2.94\times10^{-34}$                 \\
		0.0154  & $3.50\times10^{-34}\pm1.27\times10^{-35}$ & $3.40\times10^{-34}$     & $6.67\times10^{-35}$                   & $2.73\times10^{-34}$                 \\
		0.0178  & $2.24\times10^{-34}\pm1.46\times10^{-35}$ & $2.27\times10^{-34}$     & $3.43\times10^{-35}$                   & $1.93\times10^{-34}$                 \\
		0.0191  & $2.42\times10^{-34}\pm1.09\times10^{-35}$ & $1.73\times10^{-34}$     & $2.36\times10^{-35}$                   & $1.50\times10^{-34}$                 \\
		0.0204  & $1.69\times10^{-34}\pm3.22\times10^{-36}$ & $1.29\times10^{-34}$     & $1.62\times10^{-35}$                   & $1.13\times10^{-34}$                 \\
		0.0215  & $1.35\times10^{-34}\pm9.53\times10^{-36}$ & $9.90\times10^{-35}$     & $1.17\times10^{-35}$                   & $8.73\times10^{-35}$                 \\
		0.0239  & $8.92\times10^{-35}\pm5.66\times10^{-36}$ & $5.32\times10^{-35}$     & $5.70\times10^{-36}$                   & $4.75\times10^{-35}$                 \\
		0.0263  & $3.70\times10^{-35}\pm7.84\times10^{-37}$ & $2.76\times10^{-35}$     & $2.76\times10^{-36}$                   & $2.48\times10^{-35}$                 \\
		0.0287  & $2.29\times10^{-35}\pm4.56\times10^{-37}$ & $1.40\times10^{-35}$     & $1.35\times10^{-36}$                   & $1.26\times10^{-35}$                 \\
		0.0311  & $8.02\times10^{-36}\pm1.63\times10^{-37}$ & $6.93\times10^{-36}$     & $6.61\times10^{-37}$                   & $6.27\times10^{-36}$                 \\
		0.0336  & $3.79\times10^{-36}\pm1.99\times10^{-37}$ & $3.29\times10^{-36}$     & $3.24\times10^{-37}$                   & $2.97\times10^{-36}$                 \\
		0.0361  & $1.63\times10^{-36}\pm7.45\times10^{-38}$ & $1.57\times10^{-36}$     & $1.69\times10^{-37}$                   & $1.40\times10^{-36}$                 \\
		0.0385  & $6.80\times10^{-37}\pm2.50\times10^{-38}$ & $7.86\times10^{-37}$     & $9.94\times10^{-38}$                   & $6.86\times10^{-37}$                 \\
		0.0409  & $3.47\times10^{-37}\pm1.84\times10^{-38}$ & $4.04\times10^{-37}$     & $6.46\times10^{-38}$                   & $3.40\times10^{-37}$                 \\
		0.0445  & $9.28\times10^{-38}\pm2.08\times10^{-38}$ & $1.70\times10^{-37}$     & $4.14\times10^{-38}$                   & $1.28\times10^{-37}$                 \\
		0.0469  & $5.24\times10^{-38}\pm2.41\times10^{-38}$ & $1.07\times10^{-37}$     & $3.42\times10^{-38}$                   & $7.32\times10^{-38}$                 \\
		0.0529  & $1.14\times10^{-37}\pm2.48\times10^{-38}$ & $5.59\times10^{-38}$     & $2.59\times10^{-38}$                   & $3.00\times10^{-38}$                 \\
		0.0590  & $4.18\times10^{-38}\pm3.55\times10^{-38}$ & $4.31\times10^{-38}$     & $2.21\times10^{-38}$                   & $2.10\times10^{-38}$                 \\
		\hline
	\end{tabular}
\end{table*}

\begin{table*}[htbp]
	\renewcommand{\arraystretch}{1.25}
	\setlength{\tabcolsep}{20pt}
	\centering
	\footnotesize

	\caption{Experimental three-body loss rate $L_3^{\mathrm{exp}}$ as a function of magnetic field $\Delta B$ at $T=1.16~\mu\mathrm{K}$,together with theoretical results $L_3^{\mathrm{th}}$ obtained from the model fit,and its two partial contributions corresponding to $|m_\ell|=1$ and $m_\ell=0$ channels.}
	\label{tab:S4}
	\begin{tabular}{|c|c|c|c|c|}
		\hline
		$B$ (G) & $L_3^{\rm exp}$ (m$^6$/s)                 & $L_3^{\rm th}$ (m$^6$/s) & $L_3^{\rm th}(|m_{\ell}|=1)$ (m$^6$/s) & $L_3^{\rm th}(m_{\ell}=0)$ (m$^6$/s) \\
		\hline
		-0.0219 & $1.88\times10^{-38}\pm3.60\times10^{-39}$ & $3.64\times10^{-38}$     & $2.83\times10^{-38}$                   & $8.07\times10^{-39}$                 \\
		-0.0146 & $6.08\times10^{-38}\pm1.14\times10^{-38}$ & $5.26\times10^{-38}$     & $4.23\times10^{-38}$                   & $1.03\times10^{-38}$                 \\
		-0.0122 & $5.75\times10^{-38}\pm4.65\times10^{-39}$ & $6.10\times10^{-38}$     & $4.97\times10^{-38}$                   & $1.13\times10^{-38}$                 \\
		-0.0098 & $6.52\times10^{-38}\pm8.44\times10^{-39}$ & $7.19\times10^{-38}$     & $5.95\times10^{-38}$                   & $1.24\times10^{-38}$                 \\
		-0.0074 & $3.55\times10^{-38}\pm3.66\times10^{-39}$ & $8.66\times10^{-38}$     & $7.28\times10^{-38}$                   & $1.38\times10^{-38}$                 \\
		-0.0050 & $1.01\times10^{-37}\pm6.01\times10^{-39}$ & $1.08\times10^{-37}$     & $9.28\times10^{-38}$                   & $1.54\times10^{-38}$                 \\
		-0.0026 & $1.15\times10^{-37}\pm1.01\times10^{-38}$ & $1.44\times10^{-37}$     & $1.27\times10^{-37}$                   & $1.74\times10^{-38}$                 \\
		-0.0002 & $1.15\times10^{-36}\pm1.15\times10^{-37}$ & $2.05\times10^{-36}$     & $2.03\times10^{-36}$                   & $1.99\times10^{-38}$                 \\
		0.0022  & $3.24\times10^{-35}\pm4.73\times10^{-36}$ & $4.24\times10^{-35}$     & $4.23\times10^{-35}$                   & $2.30\times10^{-38}$                 \\
		0.0046  & $8.10\times10^{-35}\pm2.83\times10^{-36}$ & $8.62\times10^{-35}$     & $8.62\times10^{-35}$                   & $2.71\times10^{-38}$                 \\
		0.0083  & $1.25\times10^{-34}\pm4.46\times10^{-36}$ & $1.19\times10^{-34}$     & $1.19\times10^{-34}$                   & $3.65\times10^{-38}$                 \\
		0.0107  & $1.15\times10^{-34}\pm5.26\times10^{-36}$ & $1.21\times10^{-34}$     & $1.21\times10^{-34}$                   & $4.62\times10^{-38}$                 \\
		0.0131  & $1.19\times10^{-34}\pm3.36\times10^{-36}$ & $1.14\times10^{-34}$     & $1.14\times10^{-34}$                   & $6.14\times10^{-38}$                 \\
		0.0155  & $1.35\times10^{-34}\pm6.25\times10^{-36}$ & $1.04\times10^{-34}$     & $1.01\times10^{-34}$                   & $2.25\times10^{-36}$                 \\
		0.0179  & $1.17\times10^{-34}\pm5.52\times10^{-36}$ & $1.07\times10^{-34}$     & $8.74\times10^{-35}$                   & $1.93\times10^{-35}$                 \\
		0.0203  & $1.17\times10^{-34}\pm3.55\times10^{-36}$ & $1.16\times10^{-34}$     & $7.32\times10^{-35}$                   & $4.24\times10^{-35}$                 \\
		0.0228  & $1.28\times10^{-34}\pm8.39\times10^{-36}$ & $1.16\times10^{-34}$     & $5.97\times10^{-35}$                   & $5.60\times10^{-35}$                 \\
		0.0253  & $1.08\times10^{-34}\pm2.91\times10^{-36}$ & $1.08\times10^{-34}$     & $4.77\times10^{-35}$                   & $6.07\times10^{-35}$                 \\
		0.0277  & $1.02\times10^{-34}\pm6.22\times10^{-36}$ & $9.74\times10^{-35}$     & $3.79\times10^{-35}$                   & $5.95\times10^{-35}$                 \\
		0.0301  & $7.44\times10^{-35}\pm3.51\times10^{-36}$ & $8.42\times10^{-35}$     & $2.99\times10^{-35}$                   & $5.43\times10^{-35}$                 \\
		0.0337  & $7.82\times10^{-35}\pm9.29\times10^{-36}$ & $6.46\times10^{-35}$     & $2.05\times10^{-35}$                   & $4.41\times10^{-35}$                 \\
		0.0361  & $5.73\times10^{-35}\pm1.15\times10^{-36}$ & $5.29\times10^{-35}$     & $1.59\times10^{-35}$                   & $3.70\times10^{-35}$                 \\
		0.0410  & $4.19\times10^{-35}\pm1.45\times10^{-36}$ & $3.35\times10^{-35}$     & $9.16\times10^{-36}$                   & $2.43\times10^{-35}$                 \\
		0.0458  & $2.54\times10^{-35}\pm1.08\times10^{-36}$ & $2.05\times10^{-35}$     & $5.25\times10^{-36}$                   & $1.53\times10^{-35}$                 \\
		0.0506  & $1.75\times10^{-35}\pm9.62\times10^{-37}$ & $1.22\times10^{-35}$     & $2.98\times10^{-36}$                   & $9.25\times10^{-36}$                 \\
		0.0555  & $7.62\times10^{-36}\pm1.64\times10^{-37}$ & $7.02\times10^{-36}$     & $1.65\times10^{-36}$                   & $5.37\times10^{-36}$                 \\
		0.0615  & $4.38\times10^{-36}\pm1.60\times10^{-37}$ & $3.50\times10^{-36}$     & $7.99\times10^{-37}$                   & $2.70\times10^{-36}$                 \\
		0.0664  & $2.14\times10^{-36}\pm5.51\times10^{-38}$ & $1.94\times10^{-36}$     & $4.40\times10^{-37}$                   & $1.50\times10^{-36}$                 \\
		0.0712  & $1.21\times10^{-36}\pm6.97\times10^{-38}$ & $1.10\times10^{-36}$     & $2.49\times10^{-37}$                   & $8.48\times10^{-37}$                 \\
		0.0760  & $6.25\times10^{-37}\pm2.04\times10^{-38}$ & $6.17\times10^{-37}$     & $1.43\times10^{-37}$                   & $4.74\times10^{-37}$                 \\
		0.0808  & $3.89\times10^{-37}\pm3.09\times10^{-38}$ & $3.48\times10^{-37}$     & $8.40\times10^{-38}$                   & $2.64\times10^{-37}$                 \\
		0.0869  & $1.83\times10^{-37}\pm8.19\times10^{-39}$ & $1.74\times10^{-37}$     & $4.57\times10^{-38}$                   & $1.28\times10^{-37}$                 \\
		0.0917  & $9.96\times10^{-38}\pm6.01\times10^{-39}$ & $1.03\times10^{-37}$     & $3.00\times10^{-38}$                   & $7.35\times10^{-38}$                 \\
		0.0965  & $9.64\times10^{-38}\pm7.90\times10^{-39}$ & $6.46\times10^{-38}$     & $2.10\times10^{-38}$                   & $4.36\times10^{-38}$                 \\
		0.1013  & $2.91\times10^{-38}\pm4.43\times10^{-39}$ & $4.25\times10^{-38}$     & $1.56\times10^{-38}$                   & $2.68\times10^{-38}$                 \\
		0.1062  & $5.99\times10^{-38}\pm3.58\times10^{-39}$ & $2.96\times10^{-38}$     & $1.23\times10^{-38}$                   & $1.74\times10^{-38}$                 \\
		0.1123  & $2.10\times10^{-38}\pm2.50\times10^{-39}$ & $2.08\times10^{-38}$     & $9.72\times10^{-39}$                   & $1.11\times10^{-38}$                 \\
		0.1171  & $6.16\times10^{-39}\pm4.57\times10^{-39}$ & $1.68\times10^{-38}$     & $8.40\times10^{-39}$                   & $8.37\times10^{-39}$                 \\
		0.1219  & $1.58\times10^{-38}\pm2.68\times10^{-39}$ & $1.41\times10^{-38}$     & $7.43\times10^{-39}$                   & $6.70\times10^{-39}$                 \\
		0.1268  & $5.21\times10^{-39}\pm3.97\times10^{-39}$ & $1.23\times10^{-38}$     & $6.67\times10^{-39}$                   & $5.60\times10^{-39}$                 \\
		0.1316  & $1.02\times10^{-38}\pm3.98\times10^{-39}$ & $1.10\times10^{-38}$     & $6.08\times10^{-39}$                   & $4.87\times10^{-39}$                 \\
		0.1376  & $1.39\times10^{-38}\pm3.50\times10^{-39}$ & $9.71\times10^{-39}$     & $5.48\times10^{-39}$                   & $4.23\times10^{-39}$                 \\
		\hline
	\end{tabular}
\end{table*}

\end{document}